# Spin wave nonreciprocity and magnonic band structure in thin permalloy film induced by dynamical coupling with an array of Ni stripes

M. Mruczkiewicz,[1] P. Graczyk,[2] P. Lupo, [4] A. Adeyeye,[4] G. Gubbiotti,[3] M. Krawczyk[2*]

[1]Institute of Electrical Engineering, Slovak Academy of Sciences, Dubravska cesta 9, 841 04 Bratislava, Slovakia

[2]Faculty of Physics, Adam Mickiewicz University in Poznan, Umultowska 85, 61-614 Poznan, Poland

[3]Istituto Officina dei Materiali del CNR (CNR-IOM), Sede Secondaria di Perugia, c/o Dipartimento di Fisica e Geologia, Università di Perugia, I-06123 Perugia, Italy

[4]Information Storage Materials Laboratory, Department of Electrical and Computer Engineering, National University of Singapore, 117576 Singapore.

[*]Email: krawczyk@amu.edu.pl

## Abstract

An efficient way for the control of spin wave propagation in a magnetic medium is the use of periodic patterns known as magnonic crystals (MCs). However, the fabrication of MCs especially bi-components, with periodicity on nanoscale, is a challenging task due to the requirement for sharp interfaces. An alternative method to circumvent this problem is to use homogeneous ferromagnetic film with modified periodically surrounding. The structure is also suitable for exploiting nonreciprocal properties of the surface spin waves. In this work, we demonstrate that the magnonic band structure forms in thin permalloy film due to dynamical magnetostatic coupling with Ni stripes near its surface. We show, that the band gap width can be systematically tuned by the changing interlayer thickness between film and stripes. We show also the effect of nonreciprocity, which is seen at the band gap edge shifted from the Brillouin zone boundary and also in nonreciprocal interaction of propagating spin waves in Py film with the standing spin waves in Ni stripes. Our findings open possibility for further investigation and exploitation of the nonreciprocity and band structure in magnonic devices.

## Introduction

Periodic modification of materials has already been proven to be useful for tailoring propagation of electromagnetic, elastic and spin waves (SWs). The periodicity can be introduced along one-, two-, or three-dimensions creating band structure for waves propagating along different spatial directions. Each type of artificial crystals has many similarities common for all types of waves but also has some specific properties. Differently from electromagnetic waves which can propagate in air, spin wave excitations are confined to the magnetic materials and could not escape to the substrate or to the surrounding media. The formation of the magnonic bands separated by the magnonic band gaps have been demonstrated in 1D and 2D magnonic crystals (MCs) created by the lattice of holes, lattice of stripes or groves, and also in bi-component structures.[1,2,3,4,5,6,7,8,9] Thus, pattern introduced in thin film allows for the control of the spin wave propagation at scales comparable to its wavelength which is suitable for integration with miniaturized devices.[10,11]

The dispersion relation of long-wavelength SWs is strongly anisotropic in thin saturated ferromagnetic film, showing positive and negative slopes when waves propagate in direction perpendicular or parallel to the magnetization, respectively. The former, is Damon-Eshbach (DE) wave,[12] which is a surface wave with the amplitude concentrated at the surface of the film (this wave is also called magnetostatic surface wave). The latter waves have bulk character and due to the negative group velocity they are called backward volume magnetostatic waves. In MCs with in-plane magnetic field the waves are additionally disturbed by inhomogeneous internal magnetic field.[2,13,14,15] Inhomogeneity is created by the static demagnetizing field from the edges or interfaces in the MCs. In the case of array of stripes this demagnetizing field strongly affects the backward waves, making conditions unfavorable for propagation,[16] although propagation can be partially improved by using shallow grooves instead of stripes.[17] On the other hand, the static demagnetizing fields are absent for magnetization aligned along the stripes axis making the DE geometry preferential for investigation of the band properties. In the case of 2D array of nanodots, the anisotropic dynamical coupling for propagating SWs has also been observed when the dispersion of collective SWs has been measured with wave vector parallel and perpendicular to the applied field direction. This is ascribed to the anisotropic dipolar fields and the spatial distribution of the magnetic elements.[18]

The surface character of SWs in the DE geometry has origin in magnetostatic interaction through the dynamic demagnetizing/stray field generated by oscillating magnetization.[19] The stray magnetic field and induced electric field being outside of the ferromagnetic film can be also used to control noninvasively SW propagation.[20,21,22] Indeed, the metallic overlayer in the form of homogeneous film or a lattice of metallic stripes have been experimentally demonstrated to modify the SW spectra and propagation properties,[23] and in some cases, sufficient to create magnonic band structure, potentially useful for applications.[24] The SWs acquire additional property in such a structure – nonreciprocal dispersion, i.e., different SW dispersion relation for waves propagating in the opposite directions $\omega(k) \neq \omega(-k)$.[24,25] However, there are strong limitations regarding possible miniaturization because the influence of metal with finite conductivity is related to the screening of the microwave magnetic/electric field by conducting electrons.[26] The effect is efficient when the decay of the evanescent microwave field into the metal is smaller than the metal thickness, which requires micrometer-width metallic stripes.[27]

Propagating SWs in the ferromagnetic film can be controlled also by the ferromagnetic bar deposited above a film.[28,29] The change of the magnetization orientation in the bar changes the stray magnetic field of the oscillating magnetization in the bar which couples with the SWs in the film. This mechanism was exploited to excite SWs or control the phase and amplitude of propagating SWs in ferromagnetic stripe. The idea was further extended to the array of ferromagnetic nanodots over the ferromagnetic film. Nanodots array pumped by the microwave field at the spin wave resonance frequency have been used to induce propagating SWs in thin ferromagnetic film.[30]

Very recently, analysis of thickness-modulated single (Py) and bi-component (Py/Fe) nanowires (NWs), fabricated using developed self-aligned shadow deposition technique, have shown that layering along the third dimension is very effective for controlling the characteristics of the magnonic band. In particular, it has been found that both the frequency and the spatial profile of the most intense and dispersive mode can be efficiently tuned by the presence of the thin Fe NW overlayer. In particular, by increasing the Fe thickness, one observes a substantial frequency increase, while the spatial profile of the mode gets narrowed and moves to the permalloy NW portion not covered by Fe.[31,32]

The interesting questions arise which have not yet found answer. Firstly, does the regular pattern of the ferromagnetic substrate influence on the DE mode in homogeneous thin film to create magnonic band structure which is characteristic for MCs? If so, how sensitive is a SW dispersion of the layer to the presence of the stripe pattern? And finally, does the non-reciprocity of the SW dispersion in homogeneous film exists when the periodic pattern of the ferromagnetic substrate is introduced? We address these questions in our study experimentally through Brillouin light scattering (BLS) measurements of the dispersion relation of SWs in Py film deposited on the array of Ni stripes with different separation between layer and stripes. The experimental data are supported and explained with results of numerical computations. We demonstrated, that in homogeneous Py film the magnonic band structure is formed due to dynamic dipolar coupling with the array of the Ni stripes. We also demonstrate the existence of the effect of non-reciprocity in magnonic band structure due to non-reciprocal interaction between propagating SW in Py film and standing waves in Ni stripes.

## Sample structure and methods

Periodic arrays of nanowires (NWs) of width $w$ = 275 nm, separation of $d$ = 215 nm and lattice constant $a$ = 490 nm were patterned on Si substrate over a large area (4 mm x 4 mm) using deep ultraviolet lithography at 248 nm exposure wavelength. The substrate was first coated with an anti-reflective coating (BARC) layer of thickness of 80 nm, and a positive photoresist layer on top with a thickness of 280 nm. Details of the processing steps can be found elsewhere.[33] Cr(50 nm)/Ni (30 nm) films were then deposited subsequently on the patterned substrate using electron beam deposition technique followed by lift-off process in OK73 thinner, resulting in arrays of Cr/Ni NWs embedded in the BARC. All the layers were deposited by electron beam evaporation in a chamber with a base pressure of 2 x 10$^{-8}$ Torr without breaking the vacuum. The thickness of the multilayer stack is identical to the BARC layer to ensure a flat film surface. Fig. 1(a) shows a representative scanning electron microscopy (SEM) image of the Cr (50nm)/Ni (30nm) NWs array embedded in BARC matrix. In the final process, the structure obtained was covered by Cr($t_{Cr}$)Ni$_{80}$Fe$_{20}$(Py, $t_{Py}$ = 30nm) film with thickness $t_{Cr}$ varied from 0 to 30 nm. For the control experiments, various continuous films were also deposited at the different stages of the fabrication. The scheme of cross-section of the structure is shown in Fig. 1(b).

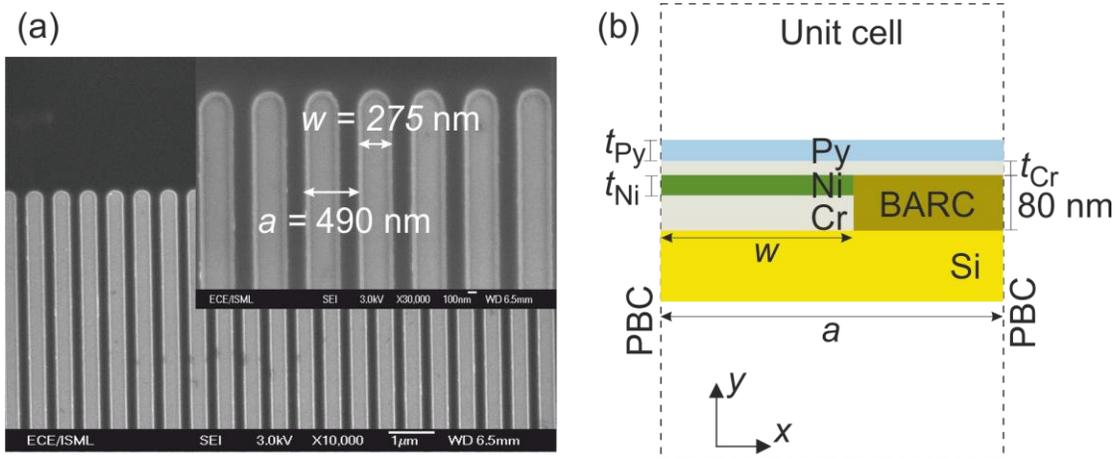

**Fig. 1.** (a) SEM image of the Ni stripes after the first stage of the fabrication process. (b) Schematic representation of the cross-sectional view of the sample (unit cell) under investigation. It is composed of a homogeneous Py film ($t_{Py}$ = 30 nm) and an array of Ni stripes ($t_{Ni}$ = 30 nm) beneath it. The SWs propagate along the $x$-axis, which is perpendicular to the magnetic field (z-direction). Periodic boundary conditions (PBC) are assumed along the $x$ axis.

Brillouin light scattering experiment have been performed by focusing about 200 mW of monochromatic laser light on top of the sample using a camera objective of numerical aperture 2 nm and focal length 50 mm. The backscattered light was analyzed in frequency by a (3+3)-tandem Fabry-Pèrot interferometer. Due to the conservation of in-plane momentum, the in-plane wave vector of SW entering in the scattering process is related to the incident angle of light by the following relation $k$ = (2π/λ)×sin(θ). The SW dispersion relation was mapped across 1.5 of the Brillouin zone (BZ) by sweeping $k$ from 0 to 20 rad/μm. A magnetic field $H$ = 500 Oe is applied in the sample plane along the easy direction of the Ni stripes and saturates magnetization in both materials. Therefore, spectra have been recorded in the DE configuration with the magnetic field perpendicular to the incidence plane of incidence of light and to the wave vector ($k$).[34]

To calculate numerically SW spectra we solve the linearized Landau-Lifshitz equation (LL) in a frequency domain with damping neglected. We assume the effective magnetic field $H_{eff}$ to be a sum of three terms: $H_{eff} = H + H_{ex} + H_{dm}$. The first term is a static bias magnetic field, $H_{ex}$ is an exchange field, and the third term $H_{dm}$ is a dynamic demagnetizing field with components along the $x$ and $y$ directions. In calculations we assume that the system is extended to infinity along the $z$-axis. In that geometry the static demagnetizing field is zero. The definition of the exchange and demagnetizing fields can be found in Ref. [**Błąd! Nie zdefiniowano zakładki.**]. We neglect magnetic anisotropy terms in the effective field, because its influence in materials investigated here is small. We solved LL equations in two-dimensional space for the unit cell (marked in Fig. 1(b) with dashed line) with the periodic boundary conditions along the $x$-axis using finite element method with COMSOL 4.3a software. From solution of the LL equation we found frequencies of the SWs for successive wavenumbers from the BZ, i.e., magnonic band structure, and spatial distribution of the SW amplitude. For more details concerning the computation method we refer to Ref. [16].

## Experimental results and Discussion

We start discussion from analysis of the dispersion relation in the reference structure, which consists of two homogeneous ferromagnetic films in direct contact. The dispersion relation measured with BLS for the bilayer composed of two 30 nm thick films of Py and Ni in direct contact are shown in Fig. 2. The BLS spectra were acquired by focusing the laser beam on the top of the Py surface (see Fig. 1 (b)). The small penetration depth of light in metals, which usually does not exceed few nm, causes that the main contribution to the collected signal comes from the uppermost Py continuous film.

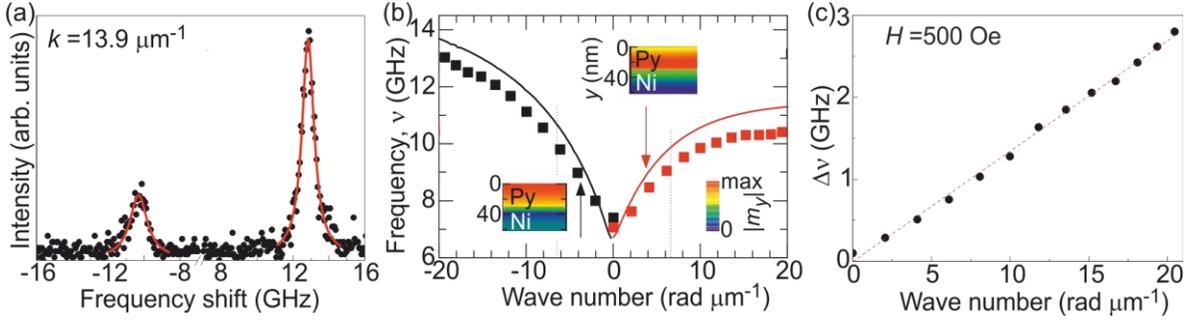

**Fig. 2.** Measured BLS spectrum (points) and fitted Lorentzian curve of the peaks on the Stokes and anti-Stokes side of the spectrum at $k$ = 13.9 rad/μm in the bilayer composed of Py and Ni films. (b) Dispersion of SWs obtained in BLS measurements. The black (red) lines and square dots show the points collected from the anti-Stokes (Stokes) side of the spectra, i.e., from the waves propagating in the +$x$ and −$x$ direction, respectively. The dots are measured data, lines refer to the calculations. In the insets the spin wave amplitude for two waves propagating in opposite direction is shown. (c) The difference in frequencies between SWs propagating in the opposite directions obtained from the Stokes and anti-Stokes sides of BLS spectra in Fig. 2(b) for the external magnetic field 500 Oe.

In Fig. 2(a) the BLS spectra for $k$ = 13.9 rad μm$^{-1}$ shows significantly different frequencies for peaks from the Stokes (negative frequency shift) and anti-Stokes (positive frequency shift) side of the spectra. The points for different incidence angles of light, and thus different wavenumbers, are collected in Fig. 2(b). The black one shows the frequency from the anti-Stokes side (along +$k$ direction) and the red one shows the frequency from the Stokes side (along -$k$) of the BLS spectra. Both lines show the dispersion relation expected for the magnetostatic SW of the DE type in bilayered film.[35,36] The dispersion starts around 7 GHz at $k$ = 0, the frequency monotonously grows with increasing wavenumber but the slope depends on the sign of $k$. With increasing $k$ a difference between frequencies of waves propagating in opposite directions $\Delta\nu(k) = \nu(k) - \nu(-k)$ increases, reaching almost 3 GHz at $k$ = 20 μm$^{-1}$. The non-reciprocity increases almost linearly with increasing wave number as shown in Fig. 2(c). It is expected that with further increase of $k$ up to the exchange-dominated region, the $\Delta\nu(k)$ will deviate from linear behaviour, reaches maximum, and starts to decrease.[37]

The numerical results for a homogeneous bilayer film are shown in Fig. 2(b) with solid lines for waves propagating in the Py film to right (black) and left (red). The dispersion relation for waves propagating in Ni film lies at lower frequencies below 6 GHz, and it is not shown here, because it is not detectable in BLS. The reasonable fit to the experimental results has been found for $M_S$ = 0.45 x 10$^6$ A/m and exchange constant $A$ = 0.8 x 10$^{-11}$ J/m, for Py: $M_S$ = 0.76 x 10$^6$ A/m and $A$ = 1.1 x 10$^{-11}$ J/m (the same parameters are used in the following part for the structures with Ni stripes). When the films are assumed to be in direct contact (full exchange coupled) the non-reciprocity was too small as compared to the value measured in BLS, for any choice of $M_S$. On the other hand, full separation of Py and Ni (exchange decoupled layers) results with too large non-reciprocity, whenever the frequency level at $k$ = 0 is matched with the experimental result. Therefore, to obtain agreement with the BLS data we introduce a small separation between Py and Ni films with a layer of very small exchange coupling. A decrease of the Ni thickness of Ni by 2 nm and introduction of 2 nm thick interlayer with the exchange constant two orders smaller than in Ni: $A_{int}$ = 0.8 x $10^{-13}$ J/m, allowed to obtain reasonable match with BLS data shown in Fig. 3(b). The separation between Py and Ni with weak exchange coupling, as suggested from the fitting procedure, can have origin in the two steps fabrication process with intermediate exposition into air. After deposition of the Ni, surface oxidation can take place, making small separation between Ni and Py. Here, the oxidation is not controlled process, which can result in inhomogeneity, and still at some parts, preserving exchange coupling.

The observed nonreciprocity in the SW dispersion relation has origin in the surface character of the DE wave, which changes the surface of localization with the change of the propagation direction.[2] The surface waves localized on the opposite surface of the Py film have different surrounding. Indeed, in the insets of Fig. 2(b) we show the amplitude distribution for SW propagating to the right and left direction along the x-axis for $k$ = 5.3 $\mu m^{-1}$. The amplitude is localized close to the top and bottom (close to Ni) side of the Py film for waves propagating in $-k$ and $+k$ direction, respectively.

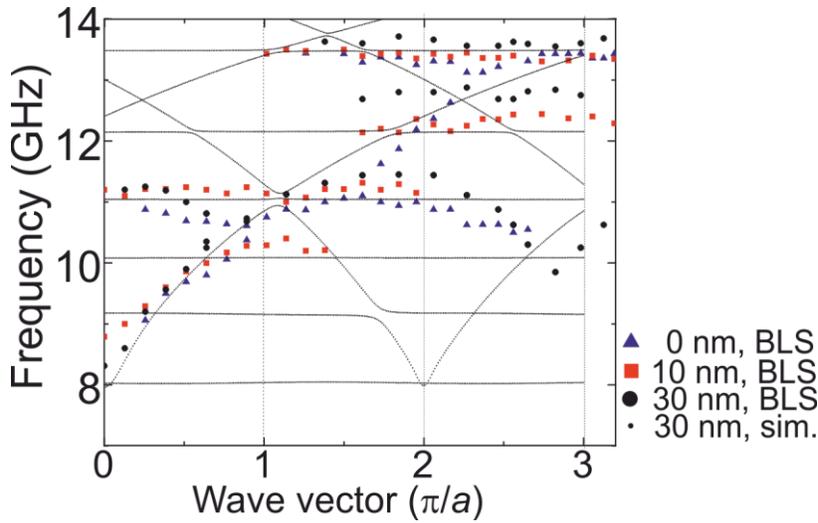

**Fig. 3.** The dispersion relation of SWs measured with BLS in Py film with the array of the Ni stripes beneath it. The data for three samples with separation $t_{Cr}$ = 0, 10 and 30 nm between Py film and Ni stripes are shown. The magnetic field is in-plane of the film and along Ni stripes, it has magnitude 500 Oe. The vertical dashed lines mark the Brillouin zone boundary. The thin lines (dense small dots) mark the numerical results for Py film and Ni stripes with separation $t_{Cr}$ = 30 nm and pinning on the Py surface.

In Fig. 3 we show the results of the BLS measurements for three samples with different values of $t_{Cr}$, in the range from 0 to 30 nm. The overall dispersion is qualitatively similar: the frequency of the lowest band starts at around 8.5 GHz at $k$=0 and it grows monotonically in the first BZ, like DE wave. Almost the same frequencies are found for peaks from Stokes and anti-Stokes side of the BLS spectra thus suggesting a negligible frequency non-reciprocity. However, closer inspection of the spectra at the border of the first BZ shows that the maximum of the first band does not appear exactly at the BZ boundary, but is shifted towards the higher wave vector value, $k \approx 1.07$ [$\pi/a$]. This feature is an indication of the non-reciprocity of SWs.[12] For higher wavenumbers (in the second BZ) a small decrease of frequency of the first at lower frequency band can be observed for the sample with $t_{Cr}$ = 10 nm. The second band in all samples has negative slope in the first BZ, and clear differences between the bands from different samples can be pointed out. The last two features are demonstration of the properties characteristic for the band structure in MCs. In the second BZ, the second band continues the trend of the dispersion of DE mode (it is clearly visible for $t_{Cr}$ = 0 nm). At higher frequencies, starting from 12 GHz, there are other modes which are dispersionless (their frequency is almost constant vs $k$). The frequency of these nondispersive lines changes with $t_{Cr}$, thus pointing at their dependence on the dipolar interaction between Py film and Ni stripes. The independence on the wavenumber can suggests the origin in the standing wave excitations resonating through the width of the Ni stripes and which are visible, even if light does not reach the Ni stripes. This can happen because of the dynamic coupling of Ni excitations to the Py film, which is confirmed in the computed field distributions.

The dispersion for Py film with Ni stripes shown in Fig. 3 is distinctly different from the dispersion for homogeneous bi-layer [Fig. 2(b)]. That points at important influence of the pattern of the Ni film on the propagating SWs in the Py film. To have deeper insight to these effects, we perform calculations for Py film with Ni stripes taking the same parameters as in a bilayer structure (with 2 nm thick layer on the top of Ni stirpes with small $A_{int}$). However, the result did not match with the BLS measurements shown in Fig. 2. The calculated dispersion was at significantly lower frequencies (it starts at 7 GHz) than in BLS measurements. The change of the magnetization saturation alone does not allow to fit with the experimental data. We need to introduce further modification to the model.

During the fabrication process (i.e., deposition of Cr and Py layers on the already patterned sample composed of BARC and Ni) the appearance of surface roughness is unavoidable. Roughness changes the conditions for magnetization dynamics locally at the interface, which effectively can change a magnetic anisotropy on the Py surface.[38,39] Let's assume, that at the bottom surface of Py film (at the interface with Cr/Ni) small surface magnetic anisotropy is induced, which results in pinning of the magnetization dynamics. To test the influence of the surface anisotropy, we performed computations with the Rado-Weertman boundary condition[40] superimposed on the dynamic part of the magnetization vector at the bottom surface of Py film (Fig. 1b):

$$pm_i + \frac{\partial m_i}{\partial x} = 0,$$

with $p$ being a pinning parameter directly connected with the surface anisotropy energy,[41] and $i$ indexes dynamical components of the magnetization vector ($i = x, y$). The use of $p$ = 0.043 nm$^{-1}$ (this corresponds to the surface anisotropy energy $4.7 \times 10^{-4}$ J/m$^2$ at the surface of Py film[42]) allows to match reasonable well the results of calculations with the BLS data, keeping parameters the same as in simulations of the bilayer structure. The results of simulations for $t_{Cr}$ = 30 nm are compared with the experimental dispersion relation in Fig. 3. We fixed value of $p$ for further investigations.

The magnonic band structure calculated for three different spacing between Py film and Ni stripes are collected in Fig. 4(a). We can distinguish two types of bands, those with nondispersive (or with weak dispersion) and with dispersive character. The amplitude and phase distribution of the low frequency excitations from the center of the BZ are shown in Fig. 4 (b-f). From their analysis we can indicate, that the nondispersive bands [(b), (c), (e)] are standing SWs in the Ni stripes, which are quantized along the nanowire width, with some forced magnetization oscillations of low amplitude visible also in the Py film. The band (d), which has finite slope at $k = 0$, is a DE mode of the Py film (Fig. 4(d)). The decrease of the SW amplitude in Fig. 4(d) at the surface close to Ni stripe for $k = 0$ is due to the pinning. The DE wave is the only excitation measured in the BLS (Fig. 3). The frequencies of DE band for different $t_{Cr}$ almost overlap in the 1st BZ. With increasing $k$ the difference between frequencies increases, but only for bands with positive slope (the band marked with f in Fig. 4a), i.e. with $+k$ propagation. In the second BZ at $k = 2\pi/a$ the DE band with positive slope of 30 nm and 0 nm sample differs by 0.28 GHz, while at $k = 4\pi/a$ the difference is already 1.4 GHz. For wave propagating in the opposite direction (the band marked with g) the influence of $t_{Cr}$ is small. This is according with the localization of the DE wave amplitude found in the bilayered structure in Fig. 2: waves propagating in Py along $+k$ and $-k$ have amplitude concentrated at the surface near and far from the Ni. Obviously, in the limit $t_{Cr} \to \infty$ the dispersion relation will tend to DE dispersion of the thin Py film, which is close to the dispersion for $-k$.

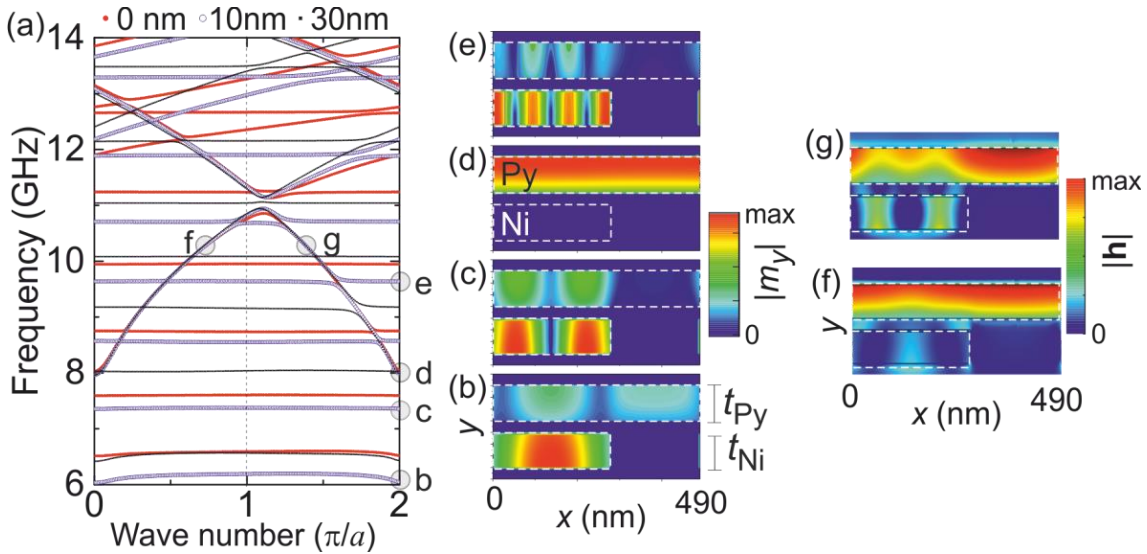

**Fig. 4**. (a) Magnonic band structure computed for the Py film with the array of Ni stripes separated by 0, 10 and 30 nm of Cr. The vertical dashed line points at the first Brillouin zone border. The empty circles labeled b – e indicate the modes of the sample with $t_{Cr} = 10$ nm from the second BZ, equivalent to the BZ center. Their SW amplitude space distributions are shown in figures (b)–(e). (f) and (g) The magnitude of stray dynamic magnetostatic field for DE mode with the same frequency (10.3 GHz) but opposite slopes, equivalent to the waves propagating in opposite directions.

The main feature of the band structure related to the periodicity is a magnonic band gap, found between 10.85 and 11.32 GHz at $k = 1.11$ $[\pi/a]$ for $t_{Cr} = 0$. The shift of the boundaries from the BZ boundary is in agreement with the BLS data. Increasing separation $t_{Cr}$ results in decrease of the gap width, accompanied with the shift of the maximum (minimum) of the first (second) band towards 1st BZ border. Both properties are shown in Fig. 5(a). The last property is a consequence of the non-reciprocity of DE mode propagation, strongly visible in the spectra of the reference bilayered structure in Fig. 2.

There is another interesting feature in the magnonic band structure shown in Fig. 4(a) related also to the non-reciprocity. Frequencies of the nondispersive bands depend on $t_{Cr}$, and there are 4 such bands in frequencies up to the magnonic band gap. Some of them interact with the DE mode resulting in the anti-crossing of the DE and horizontal bands with a significant band gap between them. Interestingly, the anti-crossing is significant only in the second BZ or alternatively for waves propagating in the -$k$ direction (due to equivalence $k' = k - G$, where $G = 2\pi/a$ is the reciprocal lattice vector). There is no (or very weak) interaction between the DE wave propagating in +$k$ direction and standing waves in Ni. This creates good opportunity for utilization of nonreciprocity induced by Ni stripes. For instance, in sample $t_{Cr}$ = 30 nm at 9.18 GHz the Py film will work as circulator or isolator for SW at wavelength around 3 μm (see the area marked with the blue dashed line in Fig. 5(b)). However, due to low intensity in the BLS spectra in the 2nd BZ the anti-crossing has not been detected experimentally. Further investigations are required to exploit this nonreciprocal effect.

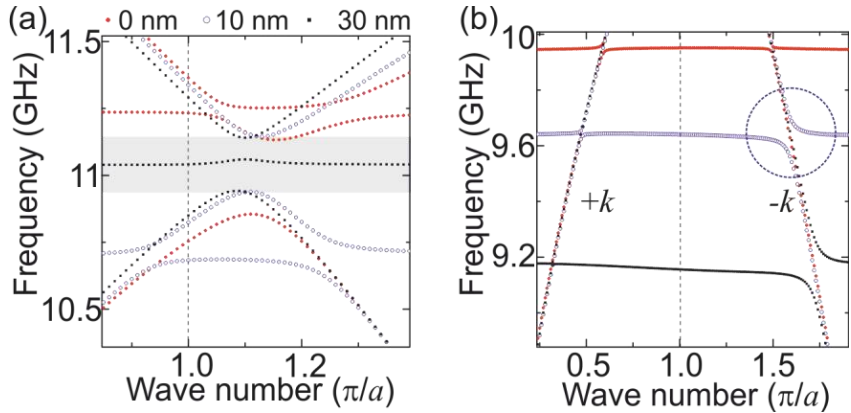

**Fig. 5.** Enlarged view of the dispersion relation from Fig. 4(a) around the magnonic band gap (marked with a gray bar for sample with $t_{Cr}$ = 30 nm) (a), and around the anti-crossings (dotted circle) of the DE band with the mode localized in Ni stripes (b).

## Summary

We have investigated experimentally with BLS and numerically with FEM method the SW dynamics in the bilayered system composed of uniformly magnetized Py thin film and array of Ni stripes beneath Py separated by a Cr spacer. We have shown that the dynamical magnetostatic coupling between the standing SW excitations in the stripes and propagating SW in the ferromagnetic film allows to form magnonic band structure with the band gaps for the SWs propagating in homogeneous Py film. This proves the possibility of controlling the band properties of SWs in homogeneous film without its patterning. We found that the spacing between the stripes and the film, if it is in the range of tens of nm, can be used to tune width of the magnonic band gap. We have also shown, that in this system the nonreciprocal effects exist. The non-reciprocity is manifested in shifting of the magnonic band gap edges from the Brillouin zone border but also in interaction between standing waves in Ni stripes and propagating SWs in Py film, which depends on the direction of the propagation.

# ACKNOWLEDGMENTS

The research has received partial funding from the People Programme (Marie Curie Actions) European Union's Seventh Framework Programme under REA Grant Agreement No. 609427 (Project WEST: 1244/02/01), the Slovak Academy of Sciences, the European Union Horizon 2020 Research and Innovation Programme under Marie Sklodowska-Curie Grant Agreement No. 644348 (MagIC) and National Science Center of Poland project no UMO-2012. AOA was supported by the National Research Foundation, the Prime Minister's Office, Singapore, under its Competitive Research Programme (CRP award no. NRF-CRP10-2012-03).